# Metabolomic response of osteosarcoma cells to nanographene oxide-mediated hyperthermia

Mónica Cicuéndez[1,2,*], Joana Flores[1], Helena Oliveira[1,3], M.Teresa Portolés[4], María Vallet-Regí[5,6], Mercedes Vila[2,+], Iola F. Duarte[1,*,+]

[1]CICECO-Aveiro Institute of Materials, Department of Chemistry, University of Aveiro (UA), Aveiro, Portugal.

[2]NRG-TEMA, Department of Mechanical Engineering, University of Aveiro (UA), Portugal.

[3]CESAM, Department of Biology, University of Aveiro (UA), Aveiro, Portugal.

[4]Department of Biochemistry and Molecular Biology I, Faculty of ~~Sciencies~~ Chemistry, ~~University~~ Universidad Complutense ~~of~~ de Madrid, Instituto de Investigación Sanitaria del Hospital Clínico San Carlos (IdISSC), Ciudad Universitaria s/n, 28040-Madrid, Spain.

[5]Department of Inorganic and Bioinorganic Chemistry, Faculty of Pharmacy, University Complutense of Madrid, Instituto de Investigación Sanitaria Hospital 12 de Octubre i+12, Ciudad Universitaria s/n, 28040-Madrid, Spain.

[6]Networking Research Center on Bioengineering, Biomaterials and Nanomedicine, CIBER-BBN, Spain.

[*]corresponding authors: ioladuarte@ua.pt; mcicuendez@ua.pt.

[+]these authors contributed equally to this work.

# ABSTRACT

Nanographene oxide (nGO)-mediated hyperthermia has been increasingly investigated as a localised, minimally invasive anticancer therapeutic approach. Near InfraRed (NIR) light irradiation for inducing hyperthermia is particularly attractive, because biological systems mostly lack chromophores that absorb in this spectral window, facilitating the selective heating and destruction of cells which have internalized the NIR absorbing-nanomaterials. However, little is known about biological effects accompanying nGO-mediated hyperthermiaat cellular and molecular levels.In this work, well-characterised pegylatednGOsheets with an hydrodynamic size of 300 nm were incubated with human Saos-2 osteosarcoma cells for 24h and their incorporation verified by flow cytometry and confocal microscopy. No effect on cell viability was observed after nGO incorporation by Saos-2 cells. However, a proliferation delay was observed due to the presence of nGO sheets in the cytoplasm.$^1$H NMR metabolomicswas employed to screen for changes in the metabolic profile of cells, as this could help to improve understanding of celular responses to nanomaterials and provide new endpoint markers of effect.Cells incorporating nGO sheets showed noticeable changes in 10 metabolites compared to control cells, including decreased levels of several amino acids, taurine and creatine and increased levels of phosphocholine and uridine/adenosine nucleotides. After NIR irradiation, cells showed decreases in glutamate and uridine nulceotides, together with increases in glycerophoscholine and adenosine monophosphate. Overall, this study has shown that the celular metabolome sensitively responded to nGO exposure and nGO-mediated hyperthermia and that NMR metabolomics is a powerful tool toinvestigate treatment responses.

# Introduction

The development of alternative and/or complementary therapeutic strategies that can efficiently eliminate tumor cells without causing major deleterious side effects is a crucially important goal in the fight against cancer[1-6]. Tumor destruction through nanomaterial-mediated hyperthermia has been increasingly investigated as a minimally invasive alternative to surgery, as well as to treat tumors embedded in vital regions where surgical resection is not feasible [7,8]. In this type of therapy, an energy-absorbing nanosystem localized within tumor tissuesabsorbs energy provided by an external source and convertsit intoheat, thereby inducing localized thermal destruction (above 40ºC)without affecting the surrounding tissues, which have not incorporated the nanomaterial [9-11].The use of Near Infrared (NIR) light in the 700-1100 nm rangefor inducing tumor hyperthermia is particularly attractive, because biological systems mostly lack chromophores that absorb in this spectral window, facilitating the selective heating and destruction of cells which have internalized the NIR absorbing-nanomaterials [12-14].

Several nanomaterialsare being developed for use in nanotechnology-based hyperthermia methods[15-20]. Nanosized graphene, particularly graphene oxide (nGO),represents one of the most thrilling candidates for this application, especially due to its strong NIR optical absorption ability [21-23].Moreover, nGOhas other fascinating properties which make it exquisitelysuitable for nanomedicine applications, such as excellent aqueous processability, amphiphilicity, Surface Enhanced Raman Scattering (SERS) property, and fluorescence quenching ability[24].

Nanographene oxide-mediated hyperthermia has been studied *in vitro*[25-27] and *in vivo*[28,29]. It is known that depending on the laser fluency, different temperature increments and various thermodynamic and thermo-biological responses can occur in the tissue medium [30]. However, there is still limited understanding of the biological effects accompanying nGO-mediated hyperthermia at the cellular and molecular levels. For instance, to our knowledge, no data concerning the metabolic changes produced in tumor cells by nGO-mediated hyperthermia have been described so far. The cellular metabolome (inventory of small molecules acting as substrates/products of enzyme-mediated reactions) closely reflects biochemical activity and functional status, being currently recognized as a key player in the cellular response to external stimuli such as nanomaterials[31,32]. Hence, the ability to comprehensively describe multiple metabolite changes (in a single analytical run) is expected to provide improved understanding of biological processes, compared to conventional single endpoint readouts.

In the present study, well-characterised pegylated nGO sheets (size of 300 nm) were incubated with human Saos-2 osteosarcoma cells for 24h and their internalization verfied, along with the impact on cell viability and proliferation. Then, $^1$H NMR analysis of lysed cell pellets was employed to assess the changes in the celular metabolome induced by nGO incorporation and subsequent laser irradiation (1.5 W/cm$^2$ power, 5 min).This approach is expected tohelp improving current understanding of cellular responses to nGO-mediated hyperthermia and to provide potential endpoint markers of effect.

## Materials and methods

### Synthesis and characterization of pegylated nGO sheets

Pegylated nGOsheetshave been obtainedfrom exfoliation of high-purity graphite in an acidic médium by a modified Hummers method, as previously reported [33]. The resulting nGO suspension was dialyzed until pH 7, and activated with chloroacetic acid (Cl–CH$_2$–COOH) under strongly basic conditions (NaOH) to promote −COOH groups at the surface. Then, it was functionalized by covalent bonding (diimide activation by adding EDAC:1-Ethyl-3-(3-dimethylaminopropyl) carbodiimide hydrochloride) with the non-toxic and non-immunogenic polymer Poly(Ethylene Glycolamine)(PEG) to improve coloidal stability and minimize immunogenicity. Finally, nGO sheets linked to PEG were marked with the amine reactive dye fluoresceinisothiocyanate (FITC) covalently bonded to the PEG. The resulting samples were analyzed by Atomic Force Microscopy AFM *multimode* Nanoscope III A (Bruker). Dynamic lightscattering (DLS) measurements were alsoperformed in pH 5 solutions in a Zetasizer Nano series instrumentequipped with a 633 nm "red" laser from Malvern Instruments, withreproducibility being verified by collection and comparison ofsequential measurements. Z-average sizes of three sequential measurementswere collected at room temperature (RT) and analyzed.X-Ray Photoelectron Spectroscopy(XPS) spectra were acquired in a Ultra High Vacuum (UHV) system with a base pressure of $2 \times 10^{-10}$ mbar.nGO-PEG powder was dispersed in MQ $H_2O$ and drop coated on a Si wafer. The system is equipped with a hemispherical electron energy analyzer (SPECS Phoibos 150), a delay-line detector and a monochromatic AlKα (1486.74 eV) X-ray source. High resolution spectra were recorded at normal emission take-off angle and with a pass-energy of 20 eV, which provides an overall instrumental peak broadening of 0.5 eV. The XPS spectra were calibrated in binding energy by referencing to the first component of the C 1s spectrum to 284.8 eV.

**Cell culture for nGO incorporation and cell proliferation studies**

Human Saos-2osteoblasts were seeded at a density of $10^5$ cells/mL in DMEM culture medium supplemented with 10%FBS, 1 mM L-glutamine, penicillin, streptomycin, under a 5% $CO_2$ atmosphere and at 37ºC for 24h. After this time, 75 µg/mL of nGO material were added to the medium and the cells were cultured for 24h in contact with nGO solution. Then, the attached cellswere harvested with 0.25% trypsin-EDTA and counted with a Neubauer hemocytometer, using trypan blue for assessing viability. For assessing nGO cellular incorporation, the fluorescence of nGO was excited at 488 nm and measured with a 530/30band pass filter in a FACScalibur Becton Dickinson flow cytometer. Theconditions for data acquisition and flow cytometric analysis were established usingnegative and positive controls with the CellQuest Program of Becton Dickinson andthese conditions were maintained during all the experiments. Each experiment wascarried out three times and single representative experiments are displayed. Forstatistical significance, at least $10^4$ cells were analyzed in each sample and the mean ofthe fluorescence emitted by these single cells was used.

**Confocal microscopy studies**

Cells were seeded on glass coverslips and cultured in the presence of pegylated nGOmaterial for 24h, fixed with 3.7% paraformaldehyde in PBS, permeabilized with0.1% Triton X-100 and preincubated with PBS containing 1% BSA. Then, cells wereincubated for 20 min with rhodamine-phalloidin (1:40), stained with 40-6-diamidino-20-phenylindole (DAPI, $3\times10^{-6}$M in PBS) and examined using a LeicaSP2 Confocal Laser Scanning Microscope. Rhodamine fluorescence was excited at540 nm and measured at 565 nm. DAPI fluorescence was excited at 405 nm andmeasured at 420-480 nm.

**NIR laser irradiation**

NIR radiation was provided by a high-power (30 W) diodelaser (LASING S.A.) emitting in 808 nm, giving a circularirradiation area of diameter 3 cm (fluency: 4 J cm$^{-2}$).The module allows irradiation of culture plates in a sterileenvironment.Irradiation was performed after nGO–cell uptake which, based on previous results, was considered complete after24h of Saos-2 exposure to nGO solution. Laser irradiation was performed at 1.5 W/cm$^2$ power during 5 min, based on previously reported results [30].Afterwards, cells were harvested using 0.25% trypsin–EDTA, washed with deuterated PBS (pH 7.4), centrifuged (1000g, 6 min, 4ºC), re-suspended in 40 µL PBS/D$_2$O, and stored at -80ºC until Nuclear Magnetic Resonance (NMR) analysis.

**NMR analysis**

Thawed cells were mechanically lysed by a three fold cycle of nitrogen freezing and sonication [34], and 50 µl of each sample were transferred into a High Resolution Magic Angle Spinning (HR-MAS) rotor. NMR spectra were acquired on a Bruker Avance DRX-500 spectrometer operating at 500.13 MHz for $^1$H observation, at 277 K, using a 4 mm HRMAS probe, in which the rotor containing the sample was spun at 4 kHz.To attenuate broad signals of macromolecules and improve the detection of small metabolites, the T2-edited Carr-Purcell-Meiboom-Gill spin-echo pulse sequence with water presaturation ('cpmgpr1d' in Bruker library) was employed, using a total spin echo time of 60 ms (n = 150, τ = 200 µs). Each 1D spectrum was acquired with 2048 transients, 32 k data points, a spectral width of 6510.42 Hz, a relaxation delay of 2 s, and an acquisition time of 2.5 s. Spectra were processed with a line broadening of 0.3 Hz and zero-filling factor of 2, manually phased and baseline corrected. Chemical shifts were referenced internally to the alanine signal at δ 1.48 ppm.Spectral assignment was based on 2D total correlation spectroscopy (TOCSY) and heteronuclear single quantum coherence (HSQC) spectra and consultation of spectral databases, such as the Bruker Biorefcode database and the human metabolome HMDB database [35].

**Multivariate analysis and integration of NMR data**

Data matrices of total area-normalized spectra were built in Amix-viewer (version 3.9.14, BrukerBiospin, Rheinstetten) and multivariate analysis (MVA) was performed in SIMCA-P11.5 (Umetrics, Sweden). Principal component analysis (PCA) was followed by partial least square discriminant analysis (PLS-DA), whereby a 7-fold internal cross-validation was applied to assess the explained variance (R2) and predictive power ($Q^2$). The corresponding loadings were obtained by multiplying the loading weight (w) by the standard deviation of each variable, and were color-coded according to variable importance to the projection (VIP).

To assess the magnitude of metabolite variations, selected spectral peaks were integrated in Amix-viewer, and normalized by the total area of each spectrum (excluding the suppressed water signal). For each metabolite, the magnitude of variation in nGO-treated/laser irradiatedsamples relatively to controls was assessed by calculating the effect size (ES), adjusted for small sample numbers, and respective standard error, according to the equations provided in the literature [36]. The metabolic changes with absolute ES larger than 0.8 (and with standard error < ES) were represented graphically and the two-sample t-test(95% confidence level) was used to assess statistical significance of variations.

**Statistics**

Data are expressed as means ± standard deviations of a representative of threeexperiments carried out in triplicate. Statistical analysis was performed using theStatistical Package for the Social Sciences (SPSS) version 19 software. Statisticalcomparisons were made by analysis of variance (ANOVA). Scheffé test was used for*post hoc* evaluations of differences among groups. In all of the statistical evaluations,$p < 0.05$ was considered as statistically significant.

## Results and discussion

### Synthesis and characterization of pegylated nGO sheets

Nanographene oxide (nGO) sheets, successfully obtained by the modified Hummers' method, were characterized by AFM,XPS, zeta-potential (ζ) and DLS particle sizeanalysis. The chemical exfoliation ofgraphite in aqueous media,followed by chemical activation and size separation bycentrifugation, resulted in pegylated nGO sheets havingan averagethickness of around 9 nm, corresponding to 13-14 layers (**Figure 1A**). Moreover, these nGO sheets had a hydrodynamic size distribution in the 200-550 nm range, with the largestnumber of nanosheets presenting sizes around 300 nm, asshown by DLS data (**Figure 1B**).Pegylated nGO sheetswere also characterized by XPS to study the elemental composition and chemical environment of the elements present within their surface.**Figure 1C** shows a fitted C 1s spectrum, composed by four components centered at: 284.8 eV (C-C/C=C), 286.3 eV (C-O/C-N), 287.7 eV (C=O) and 288.9 eV (O-C=O/N-C=O). In general, the treatment carried out to promote the carboxylation of GO from surface hydroxyl and epoxide groups diminishes the intensity of the peak of the C 1s spectrum associated to C-O groups (at 286.3 eV). Thus, the high intensity of this peak after the pegylation process indicates successful incorporation of PEG-diamine. Finally, zeta-potential analysis was carried out to evaluate the surface charge of the nGO sheets.

According to previously published data [37], the zeta-potential of chemically activatednGO sheets with -COOH surface groups is -37.8±8 mV. After the pegylation process, the zeta-potential value changed slightly to -32±6 mV, which confirms theexistence of the positive amino ended branches.

**Cell uptake of pegylated nGOsheets and cell proliferation studies**

Incorporation of the pegylated nGO sheetsand its effects on the proliferation and viability of human osteosarcomaSaos-2 cells were evaluated after 24h of incubation with 75 µg/mL of nGO. The internalization of pegylated nGO was confirmed by flow cytometry, after adding trypan blue to quench thefluorescence produced by the nGO adsorbed on theoutersurface of cells (**Figure 2A**, orange fluorescence profile) [38]. As it can be observed, similar fluorescence profiles were obtained withand without trypan blue (orange and blue profiles, respectively), indicating that pegylated nGO sheets with300 nm size and a surface charge of -32 mV were completely internalized by human Saos-2 osteoblasts after 24h of treatment.Confocal microscopy wasadditionally carried out to evaluate the Saos-2 morphology in the presenceof pegylated nGO sheets and to study its intracellular localization. **Figure 2B**and **Figure 2C**show the human Saos-2 osteoblasts morphology after 24h of treatment without and with nGO sheets, respectively. The results highlight that these pegylated nGO sheets have been incorporatedby Saos-2 osteoblasts and were dispersed throughout the cytoplasm, as indicated by arrows.These results are in agreement with those published by other authors, where most of the GO that enteredcells distributed in the cytoplasm with very few particles being present in the nucleus [39,40].As for the uptake mechanism, although this was not assessed in the present study, it has been previsouly reported that Saos-2 cells take up these pegylated nGO sheets mainly by macropinocytosis, although microtubule-dependent pathways may also be involved [41].Moreover, Chartterjee and co-wokers demonstrated that GO enters the cell through clathrin-mediated endocytosis as well as macropinocytosis [42].

The results obtained for viability and proliferation of human Saos-2 osteoblasts upon 24h nGO treatment are shown in **Figure 3**. High percentages of viability were obtained in all experimental conditions (97 and 93% in control and nGO-treated cells, respectively) and no significant differences were observed (**Figure 3A**). With respect to the effect of nGO treatment on Saos proliferation (**Figure 3B**), a significant delay in cell proliferation was observed, which is in agreement with previous results [43]. The high cell viability observed for nGO-exposed cells indicates that the plasma membrane integrity was highly preserved during the nano-bio interaction and subsequentnGO internalization by Saos-2 cells. This is in contrast with other studies where graphene and graphene-based nanomaterials were shown to alter the dynamics and integrity of the plasma membraneduring their internalization, inducing cell death[44]. Some authors suggested that the serious membrane disruption could be attributed to the strong electrostaticinteractions between the graphene surface and the lipid bilayer of the cell membrane [45]. Others revealed that certain types of graphene, such as pristine graphene, could impair cell membrane integrity by regulation of membrane- and cytoskeleton-associated genes [46]. In the present study, the absence of membrane damage is likely due to surface functionalization of the nGO sheets by the non-toxic and non-immunogenic polymer PEG [47].

As for the proliferation delay observed after nGO internalization (**Figure 3B**), the results are in agreement with those published by other authors, where a cytotoxicity evaluation of graphene oxide on different cells types was carried out[48-50]. The possible mechanism involved in the Saos-2 osteoblasts proliferation delay could be related to mitogen-activated protein kinases (MAPKs). It is well established that such kinases are involved in the regulation of cell growth, proliferation, migration and apoptosis [51]. Matesanz and co-workers reported that nGO sheets were localized on F-actin filaments of Saos-2 osteoblasts after their internalization, and thus altered cell cycle in a cytoskeleton-dependent manner [43]. Moreover, Tian X,*et al* have recently published that nGOsheets retard cellular migration via disruption of actin cytoskeleton [52]. Therefore, considering that proliferation is dependent on the cell-cycle progress, it maybe suggested that the uptake of our graphene oxide nanosheets could induce cell-cycle alterations, which would cause the proliferation delay observed.

**Metabolic response to nGO incorporation and NIR laser irradiation**

Themetabolic composition of human Saos-2 osteosarcoma cells was assessed byHRMAS $^1$H NMR analysisof lysed cell pellets. A representative cells spectrum is shown in **Figure 4**.Similarly to what has been reportedfor other tumor cell lines [53-56], the signals of several amino acids (e.g. alanine, glutamate, glycine), reduced glutathione (GSH), lactate, creatine, choline-containing compounds, taurine, myo-inisitol and uridine/adenosine nucleotides could be unambiguously identified.Phosphocholine was one of the most abundant metabolites detected, likely reflecting the high malignancy of the Saos-2 cell line, since this metabolite is considered a biomarker of tumor malignant transformation [57].Other metabolites closely related to tumor progression,such as choline and creatine (reported to be inversely expressed in later stages of tumor growth),were also clearly detected. On the other hand, glucose was absent from the Saos-2 metabolic profile, in agreement with the composition reported for osteosarcoma [58] and other tumour cell lines [53]. In non-tumorcells, the glucose flux through the glycolytic pathway is regulatedto maintain a constant concentration of adenosine triphosphate(ATP). By contrast, glucose uptake andglycolysis are known to occur nearly 10 times faster in most solid tumors compared with normal tissues [59], explaining the absence of thisimportant metabolite in tumor cells.

To assess the impact of nGO incorporation and laser irradiation on the metabolic profile of Saos-2 cells, multivariate analysis was applied to the spectra collected for the three sample groups, *i.e.* controls cells, nGO-exposed cells (nGO) and nGO-exposed and laser irradiated cells (nGO+laser). The resulting PCA scores scatter plot (**Figure 5A**) showed that control and nGO-exposed cells were reasonably clustered and separated from each other(along the PC2 axis), while the laser-irradiated samples were more scattered, reflecting higher intra-group variability. Still, by applying PLS-DA, it was possible to discriminate the three sample groups along LV1 and LV2 axes (**Figure 5B**) and to highlight the main metabolites accounting for such discrimination based on the corresponding loadings profiles (**Figures 5C** and **5D**). According to these profiles, nGO samples differed from controls in the levels of uridine nucleotides, alanine, glutamine, glycine, taurine and creatine (**Figure 5C**), while laser-irradiated samples were discriminated based on higher levels of glycerophosphocholine and AMP, together with lower levels of creatine and glutamate (**Figure 5D**). Spectral integration of individual metabolite signals was then carried out to assess the magnitude and statistical significance of the variations highlighted through multivariate analysis. The most relevant changes (absolute effect size > 0.8, as justified inref 36) are summarized in **Figure 6**. In comparison to control cells, Saos-2 cells exposed to nGO showed consistent alterations in 10 metabolites (first column of the heatmap shown in **Figure 6**), namely decreases in taurine (-44%), glutamine (-35%), alanine (-34%), creatine (-33%), glycine (-26%), methionine (-25%) and glutamate (-15%), together with increases in phosphocholine (+24%), uridine nucleotides (+44%) and AMP (+59%). As for the effects of laser irradiation, compared to nGO-exposed cells, cells irradiated after nGO incorporation showed higher levels of glycerophosphocholine (+38%) and AMP (+82%), together with decreased levels of glutamate (-34%) and uridine nucleotides (-20%) (second column of the heatmap shown in **Figure 6**).

One of the main metabolic alterations displayed by Saos-2 cells upon nGO incorporation and subsequent laser irradiation was a cumulative increase in the levels of AMP, suggesting that cellular energetic homeostasis was affected. The impact of graphene on mitochondrial function and energy production has been previously reported for other cell types [60-62]. For instance, in liver HepG2 cells, nGO sheets (400 nm) were reported to cause dysregulation of mitochondrial $Ca^{2+}$ homeostasis and a decrease in mitochondrial membrane potential (MMP), which is crucial for ATP synthesis[60]. In another work, pegylatednGO sheets in the 100-200 nm size range were shown to interact with the mitochondria of MDA-MB-231 breast tumor cells and to reduce ATP generation, possibly in relation with the down-regulation of enzymes involved in the TCA cycle and oxidative phosphorylation [61]. Moreover, in mouse alveolar macrophages, GO has been proposed to participate in redox reactions with components of the mitochondrial electron transport chain, thereby leading to impairment of ATP production [62]. In the present study, ATP was not clearly detected in the $^1$H NMR spectra, possibly due its rapid turnover. Still, the observed accumulation of AMP could be an indirect indication of energy shortage, reflecting the attempt of adenylate kinase to rescue ATP from ADP, with the concomitant production of AMP. In turn, a high level of AMP is recognized as an efficient activator of the AMP-activated protein kinase (AMPK), which is a crucial sensor of cellular energy status [63].

Upon activation, AMPK acts to restore energy homeostasis by stimulating catabolic reactions, while inhibiting ATP-consuming anabolic processes [63]. Interestingly, in the present study, the metabolic response to nGOcomprisedaconsistent decrease in the levels of several amino acids, all of which could be used as anaplerotic substrates for the tricarboxylic acid (TCA) cycle, presumably to increase energy production. Upon laser irradiation, nGO-exposed cells further decreased their glutamate levels, while the other amino acids levels (alanine, glutamine, methionine) remained similar to non-irradiated nGO-exposed cells. Moreover, nGO-incorporating cells showed decreased levels of creatine (synthetized from arginine and glycine), which could also relate to the impact of nGO on amino acid metabolism. On the other hand, phosphocholine registered a relevant increase in nGO-incorporating cells in relation to controls. As phosphocholine is central to the synthesis of the major cell membrane component phosphatidylcholine, its increment may reflect the down-regulation of phospholipid synthesis, in consonance with the putative AMPK-mediated inhibition of anabolic processes. This also agrees with the proliferation delay displayed by nGO-exposed Saos-2 cells (**Figure 3B**). Altogether, the metabolic variations described above suggest that AMPK activation may be an important player in the cellular responses to nGO and nGO-mediated hyperthermia, an hypothesis which needs to be verified in further studies.

Laser-irradiated cells further showed an increase in glycerophosphocholine, a well-known membrane breakdown product, likely reflecting heat-induced membrane damage[64]. Indeed, previous results demonstrated that nGO mediated hyperthermia under similar conditions to those applied in this study (low power and 7 min of laser exposure) caused Saos-2 osteoblasts death by mixed apoptotic/necrotic process, implying loss of the plasma membrane integrity [30]. Notably, glycerophosphocholine levels did not vary upon nGO incorporation in the absence of laser treatment, in agreement with the preservation of membrane integrity in non-irradiated nGO-incorporating cells (**Figure 3A**).

As taurine is an antioxidant metabolite [65], its decrease upon nGO incorporation could be related to nanomaterial-induced oxidative stress. However, while the generation of reactive oxygen species (ROS) has been reported to be a common response to graphene-based nanomaterials[40,42], previous results from our group have shown that pegylated nGO sheets of similar size did not increase ROS in Saos-2 cells [43]. Moreover, glutathione, another important antioxidant metabolite clearly detected in the cells profile, did not vary upon nGO incorporation, corroborating the absence of a strong oxidative stress response. Therefore, taurine variation likely relates to other possible roles of this metabolite, such as in osmoregulation, membrane stabilization, calcium homeostasis or protein phosphorylation [66].

Finally, nGO-incorporating cells showed increased levels of uridine nucleotides, which subsequently decreased after laser irradiation. The bioavailability of uridine is particularly crucial to the synthesis of RNA and biomembranes (via the formation of pyrimidine nucleotide–lipid conjugates), being also needed for the formation of UDP-sugar conjugates involved in the post-translational modification of proteins [67]. Therefore, the observed variations may possibly relate to the effects of nGO and hyperthermia on cell proliferation and/or protein glycosylation.

## Conclusions

In this work, Saos-2 osteosarcoma cells were shown to incorporate 300 nm pegylated nGO sheets without impairing cell viability, while causing a delay in cell proliferation. These results were in agreement with other studies where nGO was found to distribute in the cells cytoplasm [39,40] and to interfere with cell cycle and growth [43]. At the metabolic level, nGO produced consistent variations in 10 metabolites, mainly reflecting changes in energy homeostasis, also seen in previous studies using other cell types and different assays to assess mitochondrial function and energy production [61-63].In particular, based on increased levels of AMP and phosphocholine, together with decreases in several amino acids, it has been newly hypothesized that AMPK activation may be an important player in the cellular responses to nGO. This hypothesis should be verified in future studies, as it could provide improved mechanistic understanding of nGO mode of action at the molecular level. Notably, subsequent laser irradiation of nGO-incorporating cells produced fewer effects: changes in AMP and glutamate were amplified, the levels uridine nucleotides recovered and glycerophosphoholine increased, this latter change likely reflecting laser-induced membrane damage, thus agreeing with previous results [30]. Membrane integrity was, however, not affected upon nGO incorporation, unlike the results reported by others where non-pegylated nGO was employed [45,46]. Overall, this study has shown that cellular metabolome sensitively responded to nGO exposure and nGO-mediated hyperthermia, and that NMR metabolomics is a powerful tool in nanotoxicity studies, allowing for new hypotheses on molecular mechanisms underlying treatment responses to be generated.

## Acknowledgements

This work was developed within the scope of the project CICECO-Aveiro Institute of Materials, POCI-01-0145-FEDER-007679 (FCT Ref. UID /CTM /50011/2013), financed by national funds through theFCT/MEC and when appropriate co-financed by FEDER under the PT2020 Partnership Agreement. M.C. acknowledges the FCT financial support [Post-Doctoral Grant SFRH/BPD/101468/2014] and Operational Program Human Capital (POCH), European Union. H.O. acknowledges financial support FCT SFRH/BPD/111736/2015. M.T.P.acknowledges funding from Ministerio de Economía y Competitividad (projects MAT2013-43299-R and MAT2016-75611-R AEI/FEDER, UE). M.V.R.acknowledges funding from the European

Research Council (Advanced Grant VERDI; ERC-2015-AdG Proposal No. 694160). I.F.D. acknowledges the FCT/MCTES for a research contract under the Program 'Investigador FCT' 2014. The PortugueseNational NMR Network supported with FCT funds and Bruker BioSpin GmbH for database access are also acknowledged. Thanks also to the staff of the Centro de Citometría y Microscopía de Fluorescencia of the Universidad Complutense de Madrid (Spain) and ICTS Centro Nacional de Microscopia Electrónica (Spain) for the assistance in the flow cytometry, confocal microscopy and AFM studies, respectively.

# Figure captions

**Figure 1.** Characterization of pegylated nGO sheets. **(A)** AFM topographic image of dispersed nGO sheets. The graph represents the thickness profile obtained from the white line of the AFM image. **(B)** nGO sheetsparticle size distribution obtained by DLS. **(C)** C1s XPS spectra obtained from nGO sheets.

**Figure 2.** Cell uptake of pegylated nGO sheets by human Saos-2 cells. **(A)** Fluorescence profiles of Saos-2 osteoblasts cultured for 24h with nGO (Saos + nGO) compared to control cells in absence of nanomaterial (Saos), quenching the exterior cell surface fluorescence with trypan blue. Morphology evaluation by confocal microscopy of cultured human Saos-2 cells after 24h of treatment without **(B)** and with **(C)** nGO sheets. Cells were stained with DAPI for the visualizationof the cell nuclei in blue and rhodamine-phalloidin for the visualization of cytoplasmic F-actin filaments in red.

**Figure 3.** Effect of pegylated nGO sheets on the viability percentage **(A)** and proliferation **(B)** of human Saos-2 osteosarcoma cells after 24h of treatment.Statistical significance *$\rho<0.05$

**Figure 4.** 500 MHzHRMAS$^1$H NMR spectrum of human Saos-2 osteosarcoma cells, with some assigned metabolites indicated.AMP: adenosine monophosphate, BCAA: branched chain amino acids (valine, leucine and isoleucine), Cr: creatine, Glu: glutamate, GSH: glutathione, Lac: lactate, m-Ino: myo-inositol, PC: phosphocholine, PCr: phosphocreatine, Phe: phenylalanine, Tau: taurine,Thr: threonine, Tyr: tyrosine, UDP: uridine diphosphate.

**Figure 5.** Multivariate analysis of human Saos-2 osteosarcoma cells spectra. **(A)** PCA scores plot, **(B)** PLS-DA scores plot **(C)** PLS-DA LV2 loadings: Ct *vs.* nGO. **(D)** PLS-DA LV1 loadings: nGO *vs.* nGO + laser.

**Figure 6.** Heatmap of mainmetabolic variations after nGO incorporation by human Saos-2 osteoblasts (nGO vs. Ct) and after NIR laser irradiation at 1.5 W/cm$^2$ power during 5 min (nGO+Laser vs. nGO). Statistical significance **$\rho<0.01$.

**Figure 1**

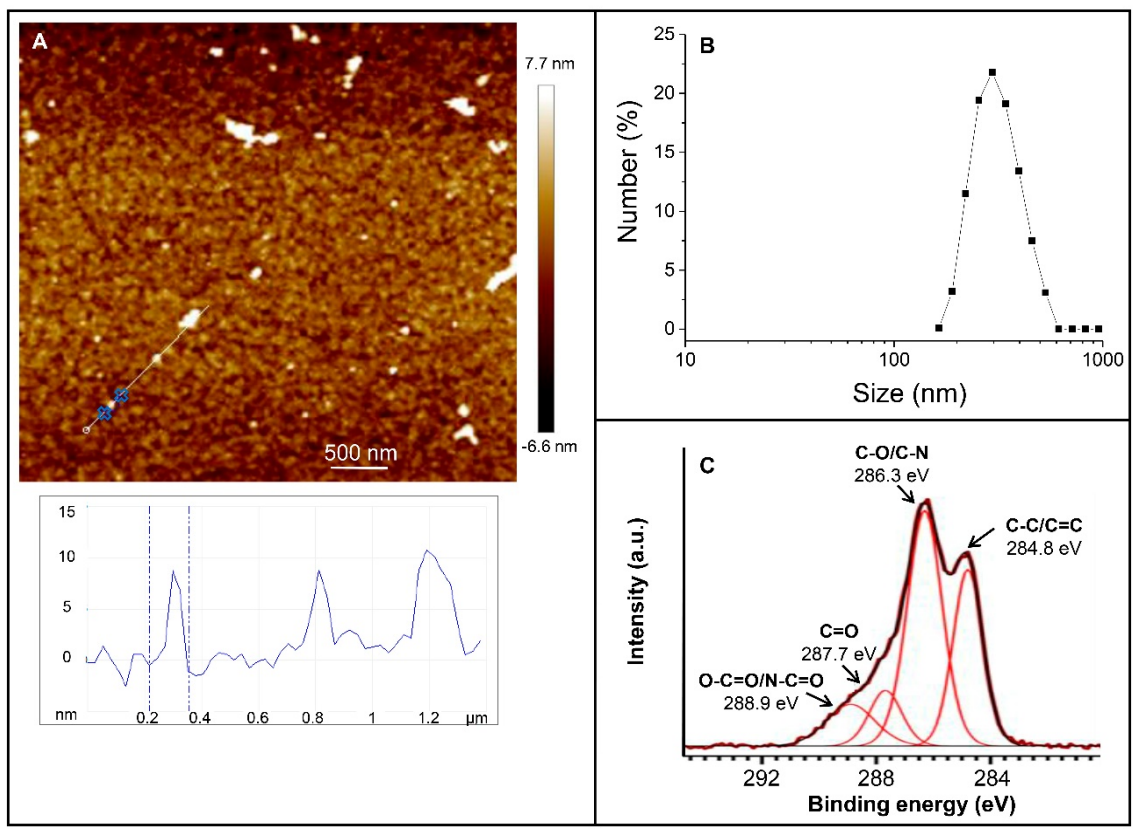

**Figure 2**

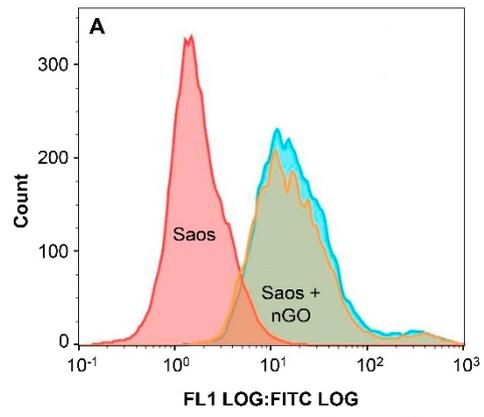

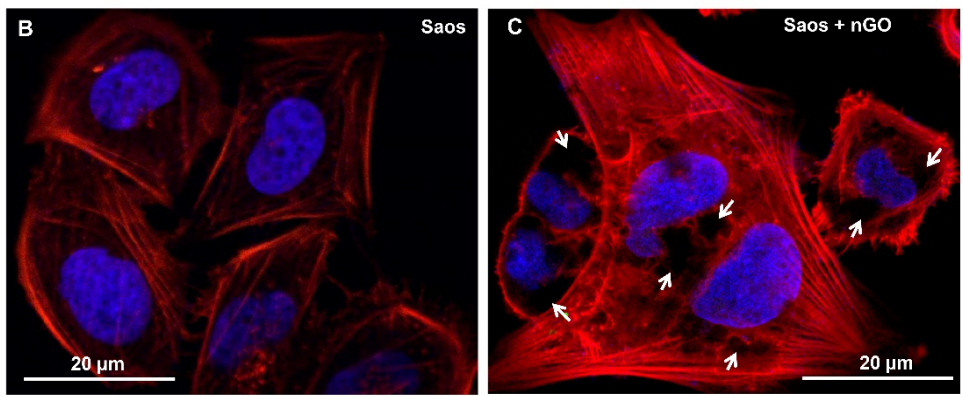

**Figure 3**

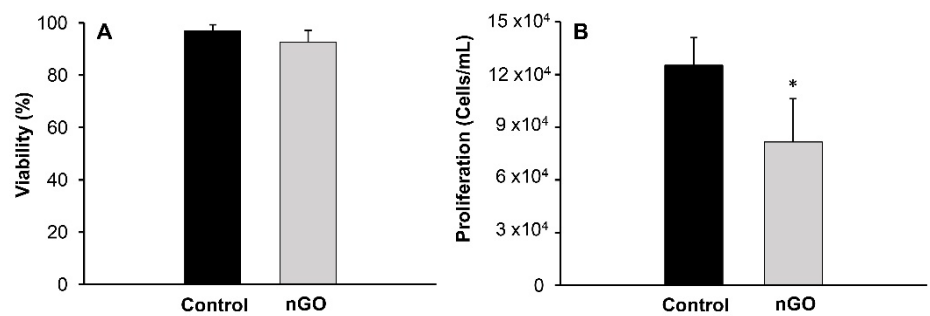

**Figure 4**

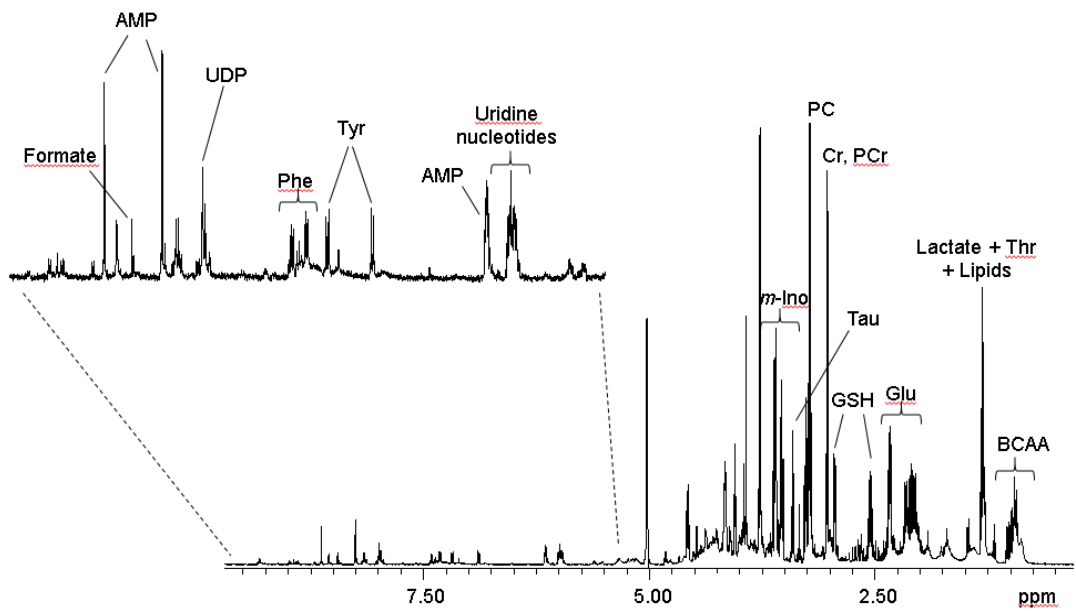

**Figure 5**

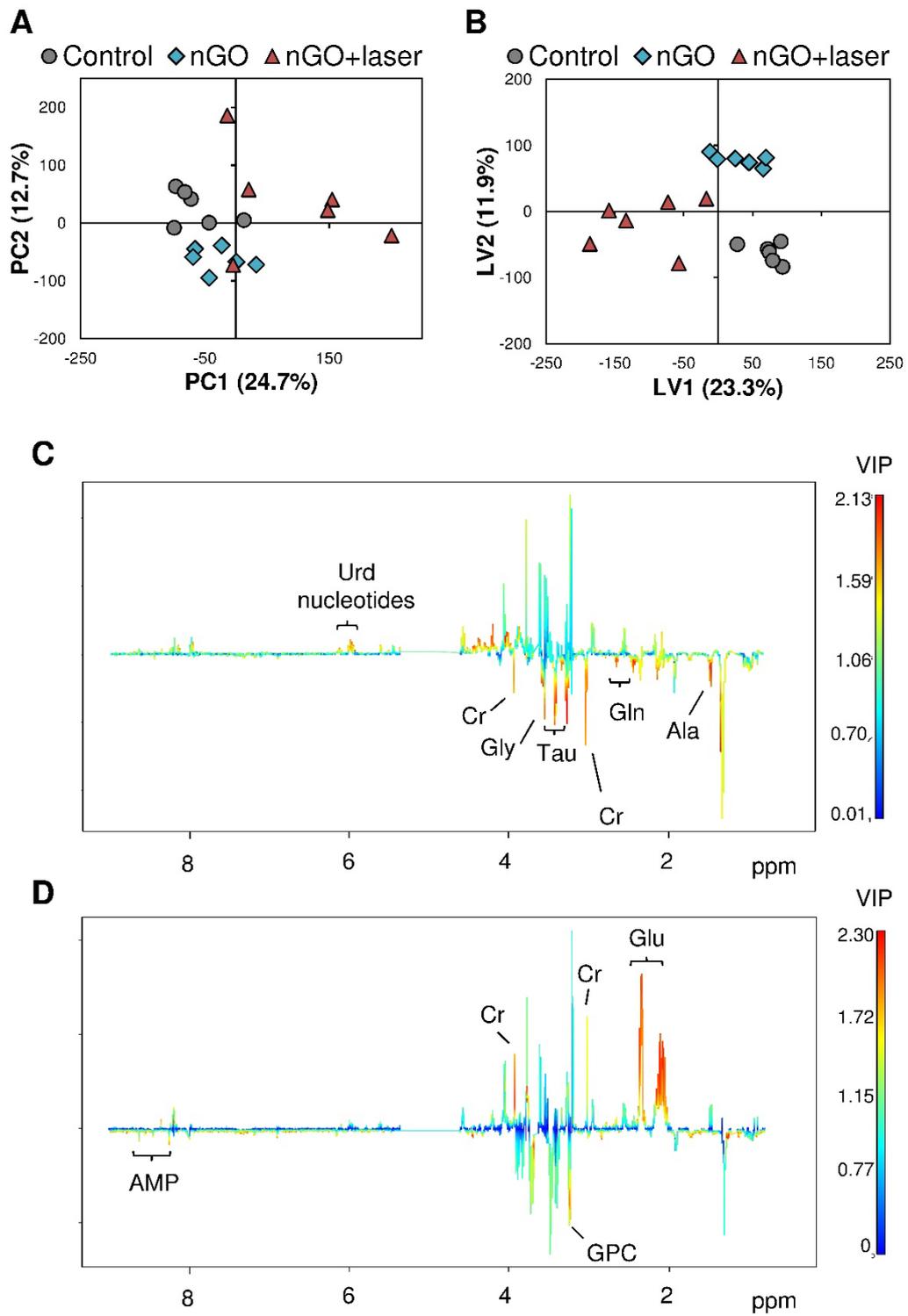

**Figure 6**

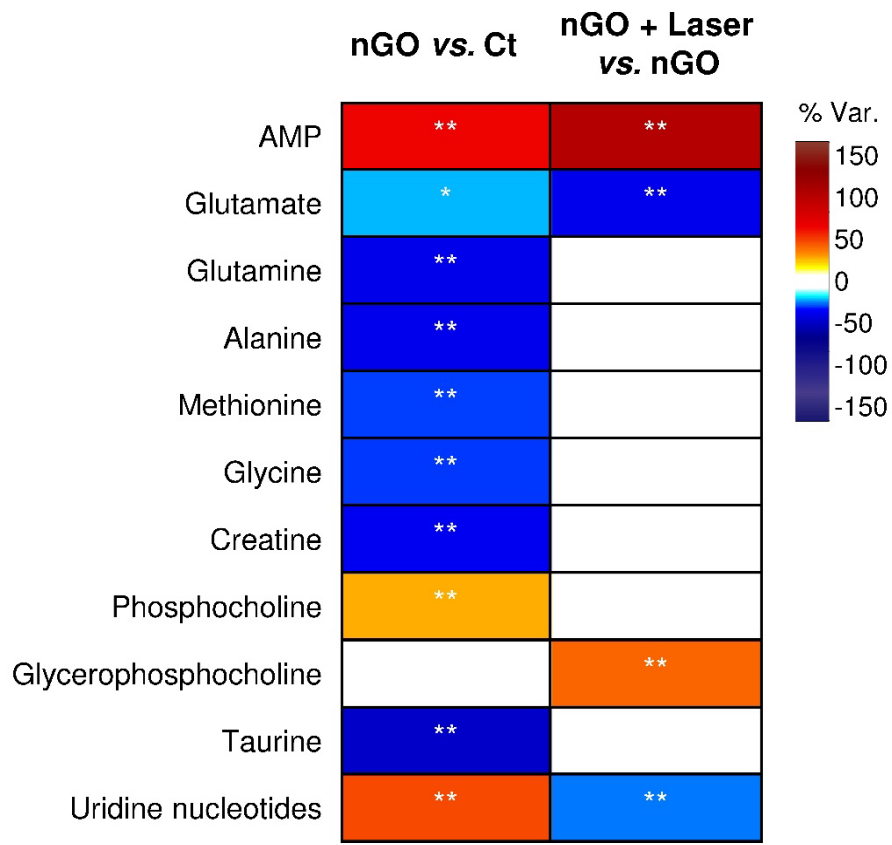

# Graphical Abstract

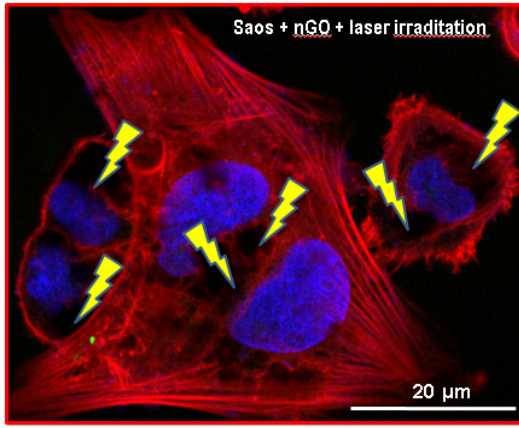